\documentclass[letterpaper]{jpconf}
\usepackage{graphicx}
\def\beq{\begin{equation}}
\def\eeq{\end{equation}}
\def\bea{\begin{eqnarray}}
\def\eea{\end{eqnarray}}

\newcommand{\cA}{{\cal A}}
\newcommand{\cAb}{{\overline{\cal A}}}
\newcommand{\cF}{{\cal F}}
\newcommand{\cFb}{{\overline{\cal F}}}
\newcommand{\cD}{{\cal D}}
\newcommand{\cDb}{{\overline{\cal D}}}
\newcommand{\cQ}{{\cal Q}}
\newcommand{\cU}{{\cal U}}
\newcommand{\cN}{{\cal N}}
 
\newcommand{\etab}{{\overline{\eta}}}
\newcommand{\psib}{{\overline{\psi}}}
\newcommand{\phib}{{\overline{\phi}}}

\def\nn{\nonumber}
\def\bec{\begin{center}}
\def\eec{\end{center}}
\begin{document}
\title{Gauge theory duals of black hole - black string transitions of gravitational theories on a circle}
\author{Simon Catterall$^1$, Anosh Joseph$^1$\footnote[2]{Speaker. PREPRINT \# SU-4252-905.} and Toby Wiseman$^{3}$}
\address{$^1$ Department of Physics, Syracuse University, Syracuse, New York, 13244 USA}
\address{$^3$ Theoretical Physics Group, Blackett Laboratory, Imperial College, London SW7 2AZ, UK}
\ead{smc@physics.syr.edu, ajoseph@physics.syr.edu, t.wiseman@imperial.ac.uk}
\begin{abstract}
We study the black hole - black string phase transitions of gravitational theories compactified on a circle using the holographic duality conjecture. The gauge theory duals of these theories are maximally supersymmetric and strongly coupled $1 + 1$ dimensional $SU(N)$ Yang-Mills theories compactified on a circle, in the large $N$ limit. We perform the strongly coupled finite temperature gauge theory calculations on a lattice, using the recently developed exact lattice supersymmetry methods based on topological twisting and orbifolding. The spatial Polyakov line serves as relevant order parameter of the confinement - deconfinement phase transitions in the gauge theory duals.
\end{abstract}
\section{Introduction}
The rise of string theory as a candidate for a unified quantum theory including gravity has generated a resurgence of interest in black holes. In an appropriate limit, string theory reduces to supergravity. There are different types of black hole solution possible in supergravity and more interestingly, there are phase transitions possible between these black solutions. Examples of such transitions include the Hawking - Page transition \cite{Hawking:1982dh} in AdS spaces and the Gregory - Laflamme transition \cite{Gregory:1993vy} in $D1$-brane systems compactified on a circle \cite{Li:1998jy}. These phase transitions are very difficult to tackle analytically in the gravitational theories. However, they have dual descriptions as thermal phase transitions in strongly coupled gauge theories. It is possible to analyze these phase transitions by studying the phase structure of their strongly coupled dual gauge theories. The dynamics of strongly coupled dual gauge theories can be studied on a lattice, using the recently developed exact lattice supersymmetry methods based on topological twisting and orbifolding \cite{Catterall:2009it}.

The holographic duality conjecture in string theory, in an appropriate limit, relates the finite temperature low energy supergravity limit of string theory to strongly coupled finite temperature supersymmetric Yang-Mills theory. For a large number $N$ of coincident $Dp$-branes in the ``decoupling" limit, the dual gauge theory is $(1 + p)$-dimensional, strongly coupled, maximally supersymmetric $SU(N)$ Yang-Mills theory taken at large $N$. 

The decoupling limit is described as follows. String theory contains $D$-branes, which are solitonic objects. When we consider the full theory in the presence of these solitons we have modes that propagate in the bulk and modes that propagate on the solitons. The modes on the soliton interact with each other and with the bulk modes. It is possible to define a limit of the full theory in which the bulk modes decouple from the modes living on the $D$-brane. This is typically a low energy limit, in which we tune the coupling constant so as to keep only the interactions among the modes living on the $D$-brane. In this limit the $D$-brane theory becomes super Yang-Mills theory (for the case $p \leq 3$).

The simplest case corresponds to $D0$-branes, where the dual theory is a finite temperature 16 supercharge Yang-Mills quantum mechanics with gauge group $SU(N)$ \cite{Itzhaki:1998dd}. (See \cite{Catterall:2007fp} for recent lattice studies of super Yang Mills quantum mechanics.) When $p=3$, we have the case of $D3$-branes and the original AdS/CFT correspondence \cite{Maldacena:1997re}. 

In this talk, we focus on the gauge theory duals of black hole - black string phase transitions due to Gregory - Laflamme instabilities developed in a system of $N$ black $D1$-branes compactified on a circle. In the gauge theory dual interpretation, these transitions correspond to thermal phase transitions in $1+1$ dimensional maximally supersymmetric $SU(N)$ Yang-Mills theory on a circle at strong coupling \cite{Aharony:2004ig}.
\section{Two dimensional super Yang-Mills and $D1$-branes}
\subsection{Maximally supersymmetric Yang-Mills on $T^2$}
We consider the $1+1$ dimensional 16 supercharge strongly coupled SYM theory with gauge group $SU(N)$. It can be obtained from the dimensional reduction of $\cN = 4$, $d=4$ SYM theory into two dimensions. The gauge theory is at strong coupling $\lambda = g^2_{YM} N \gg 1$, where $\lambda$ is the `t Hooft coupling, $g_{YM}$ is the Yang-Mills coupling and $N$ is the number of colors.

It is natural to consider the case where the spatial dimension is compactified on a circle of circumference $L$. Since we are interested in studying the thermodynamics of this gauge theory on a lattice, we consider the Euclidean version of the theory. To study the field theory at a finite temperature we compactify the temporal direction on a circle with period $\beta$. That is $t \sim t + \beta$, where $\beta = 1/T$ is the inverse temperature. Thus the two dimensional Euclidean $SU(N)$ Yang-Mills theory is compactified on a rectangular torus $T^2$.

Euclidean $1+1$ dimensional $SU(N)$ SYM on $T^2$ has two non-contractible Wilson loops (Polyakov lines), namely the Wilson loops running around the temporal (thermal) circle ($P_t$) and the spatial circle ($P_x$). The fermions are anti-periodic on the thermal circle and we impose periodic boundary conditions on the spatial circle. In this way, we can distinguish between spatial and temporal directions of the compactified theory.

The compactified theory on $T^2$ exhibits phase transitions as we change the temperature. The expectation value of the spatial Polyakov loop 
\beq
P_x = \Big{\langle} \Tr (P \exp (i \oint A_x))\Big{\rangle}~,
\eeq
where $A_x$ is the gauge field along $x$-direction, in the Euclidean theory serves as order parameter for this phase transition.

The expectation value of $P_x$ changes from zero to non-zero values as we increase the temperature. The expectation value of $P_t$ is always non-zero at all temperatures. Thus upon heating, the SYM theory undergoes a transition from a ``confined phase" ($P_x = 0$) to a ``deconfined phase" ($P_x \neq 0$).
\subsection{The gravitational theory on a circle}
The dual gravitational theory corresponding to the $(1+1)$ dimensional maximally supersymmetric $SU(N)$ Yang-Mills theory describes the near-horizon geometry of the $D1$-brane metric of type IIB string theory \cite{Itzhaki:1998dd}.

The string theory contains a collection of $N$ $D1$-branes. The decoupling limit considers keeping finite energy excitations of the $D1$-branes while taking the limit,~~$g^2_{YM} = \frac{1}{2\pi}\frac{g_s}{\alpha'} = \textrm{fixed},~~\alpha' \rightarrow 0$, where $g_s$ is the string coupling and $\alpha'$ is the slope parameter. 

Since the dual gauge theory is at finite temperature $T$, we are interested in near extremal black $D1$-brane configurations of the gravitational theory in the above limit. 
We have compactified the spatial direction of the gauge theory on a circle of circumference $L$. This implies that we should impose a periodic identification of the spatial coordinate in the dual gravitational theory. Thus we have a space that is asymptotically $R^{8,1} \times S^1$.

We may write a IIB supergravity approximation for the string frame metric and dilaton (for a charged black string winding around $S^1$) as
\bea
ds^2 &=& \alpha' \Big\{ \frac{u^3}{\sqrt{d_1 \lambda'}} \Big[ - \Big(1 - \frac{u_0^6}{u^6}\Big) \frac{d\tau^2}{L^2} + \frac{d\theta^2}{(2\pi)^2} \Big] + \frac{\sqrt{d_1 \lambda'}}{u^3 \Big(1 - \frac{u_0^6}{u^6}\Big)}du^2 + u^{-1}\sqrt{d_1 \lambda'}d\Omega^2_7\Big\}~,
\eea
\beq
e^{\phi} = 2 \pi \frac{\lambda'}{N} \sqrt{\frac{d_1 \lambda'}{u^6}}~,
\eeq
where $\lambda' = \lambda L^2$, $d_1 = 2^6 \pi^3$, and $\theta$ is the angular periodic coordinate describing the spatial $S^1$, so that $\theta \sim \theta + 2 \pi$. From this supergravity solution we may derive the thermodynamics of the black hole from the region of the solution near the horizon using standard methods. One can compute that the constant of integration $u_0$ is related to the temperature of the solution as $u_0^2 = \frac{16 \pi^{5/2}}{3} t \sqrt{\lambda'}$, where we have defined the natural dimensionless temperature $t = TL$. If one continues time to obtain a Euclidean solution, then there is a conformal boundary which is a torus $T^2$ whose cycles derive from the spatial circle and the Euclidean time circle \cite{Aharony:2004ig}.

Analysis of $\alpha'$ corrections implies that this supergravity solution near its horizon is a good approximation to the actual IIB string background provided that $t << \sqrt{\lambda'}$. The compact spatial circle means one must also ensure that winding string modes are not important corrections to the supergravity near the horizon, and this implies $1/\sqrt{\lambda'} << t$. Thus for large $\lambda' >> 1$ this supergravity solution correctly approximates the horizon region of the true string background for a range of temperatures. This solution is believed to be perturbatively stable within the supergravity approximation. 

Below dimensionless temperatures  $t \sim 1/\sqrt{\lambda'}$, the above background is strongly effected by string winding modes near the horizon which wrap over the spatial circle. We then cannot analyse the IIB string theory behaviour  simply using IIB supergravity. However, we may employ the trick of T-duality, acting on the spatial circle, to map the IIB solution to one in IIA theory which in fact can be described in IIA supergravity. One finds the following IIA solution after T-duality,
\bea
ds^2 &=& \alpha' \Big\{ - \frac{u^3}{\sqrt{d_1 \lambda'}} \Big(1 - \frac{u_0^6}{u^6}\Big) \frac{d\tau^2}{L^2}  + \frac{\sqrt{d_1 \lambda'}}{u^3} \left[ \frac{du^2}{\Big(1 - \frac{u_0^6}{u^6}\Big)} + (2 \pi)^2 d\tilde{\theta}^2 \right] + u^{-1}\sqrt{d_1 \lambda'}d\Omega^2_7\Big\}~,
\eea
\beq
e^{\phi} = (2 \pi)^2 \frac{\lambda'}{N} \left( \frac{d_1 \lambda'}{u^6} \right)^{3/4}~,
\eeq
now with $\tilde{\theta}$ as the angular coordinate on the (T-dual) spatial $S^1$, again with $\tilde{\theta} \sim \tilde{\theta} + 2 \pi$. The requirement that the supergravity solution and its perturbations are good, gives the condition $\lambda' >> 1$ and $t << 1$ from requiring $\alpha'$ and winding corrections to be small near the horizon. 

A gravity analysis now shows that in the T-dual IIA supergravity picture the solution above is perturbatively unstable to a gravity mode that breaks the translational invariance along the $\tilde{\theta}$ circle direction, and is localized near the horizon. This is an instability of the Gregory-Laflamme type. The analysis shows that it occurs below a temperature,
\beq
t_{GL} = \frac{3}{4 \sqrt{\pi}} \frac{(2 \pi a)^2}{\sqrt{\lambda}'}
\eeq
where $a$ is a constant determined numerically to be $a \approx 0.37$. We note that this is in the regime of validity of the IIA supergravity solution provided $\lambda' >> 1$.

In the T-dual IIA picture which allows us to describe the $D1$-branes using supergravity we see the instability as a perturbative graviton mode that occurs for $\lambda' >> 1$ at low temperatures $t$. Under T-duality the $D1$-branes wrapped on the $S^1$ become a density of $D0$-branes `smeared' over the $S^1$. Physically this configuration becomes unstable to clumping of the $D0$ branes at low temperature. However, the interpretation in terms of the original $D1$-branes with $\lambda' >> 1$ is as a stringy winding mode instability about the spatial $S^1$, which occurs for $t \sim 1/\sqrt{\lambda'}$ (and hence can't be described using the IIB gravity solution) and is associated with symmetry breaking on the circle.

From a thermal Euclidean perspective the AdS/CFT correspondence therefore maps (the near-horizon limit of) near extremal solutions of the full type IIB string theory compactified on a circle to thermal phases of maximally supersymmetric Yang-Mills theory on $S^1 \times S^1$.
In the gauge theory the natural interpretation of the above stringy instability is a transition from a ``confining phase'' ($P_x = 0$) to a ``deconfining phase'' ($P_x \neq 0$) as the dimensionless temperature $t$ varies. The expectation value of $P_t$ is always non-vanishing at all temperatures indicating the presence of event horizon in the dual gravitational theory, or from a Euclidean geometric perspective, the contractibility of the Euclidean time circle.

\section{Strongly coupled $2D$ super Yang-Mills on a lattice}
In recent years a new approach on how to put supersymmetric field theories on a lattice has been developed using ideas drawn from topological twisting \cite{Catterall:2003wd} and orbifolding \cite{Cohen:2003xe}. This approach indicates that a class of supersymmetric theories may be discretized by preserving just a fraction of the original supersymmetry. (See \cite{Catterall:2009it} for a review.)  

The approach mentioned above, exposes a nilpotent supersymmetry. This supersymmetry say, ${\cal Q}$ is a Lorentz scalar and it does not produce translations on the lattice. The nilpotent property allows us to rewrite the two-dimensional super Yang-Mills action as a ${\cal Q}$-exact term. This action can then easily be discretized on a lattice.
\subsection{Topological twisting}
In this talk we discuss the results derived from the lattice simulations of topologically twisted two-dimensional (Euclidean) super Yang-Mills theory with gauge group $SU(N)$. The fields are all in the adjoint representation with an anti-hermitian basis for the generators. The two-dimensional 16 supercharge theory can be obtained from dimensional reduction of the four-dimensional ${\cal N} = 4$ SYM down to two dimensions. 

We rewrite the original action of the four-dimensional theory in terms of ``twisted fields." The basic idea of twisting is not new, it goes back to Witten in his seminal paper on topological field theory \cite{Witten:1988ze}. In the case of four dimensional  ${\cal N} = 4$ SYM, the process of twisting is a decomposition of the fields in terms of representations of a twisted rotation group, which is a diagonal subgroup of the original (Euclidean) rotational symmetry $SO_{\rm Lorentz}(4)$ and an $SO_{\rm R}(4)$ subgroup of the R-symmetry of the theory. (The four-dimensional theory has $SO_{\rm R}(6)$ R-symmetry.) The twisted rotation group is:
\beq
SO(4)^\prime={\rm diag}\Big(SO_{\rm Lorentz}(4)\times SO_{\rm R}(4)\Big)~.
\eeq

We can treat the supercharges of the twisted four-dimensional ${\cal N} = 4 $ theory as a $4 \times 4$ matrix $q$. This matrix can be expanded on products of gamma matrices
\beq
q=\cQ I + \cQ_\mu \gamma_\mu + \cQ_{\mu \nu}\gamma_\mu\gamma_\nu + \ldots 
\eeq
After the twist the original supersymmetry algebra becomes a twisted algebra
\begin{eqnarray}
\cQ^2&=&0\\
\{\cQ,\cQ_\mu\}&=&p_\mu\\
&\cdots&
\end{eqnarray}
The first piece of the algebra shows that the nilpotent scalar supercharge $\cQ$ does not produce translations on the lattice. That is, we can easily write down the twisted theory on a lattice. The second piece of the algebra expresses the fact that the momentum is the $\cQ$-variation of something, which makes plausible the statement that the energy-momentum tensor and hence the entire action can be written in $\cQ$-exact form. An action written in such a $\cQ$-exact form is trivially invariant under the scalar supersymmetry provided the latter remains nilpotent under discretization.

The rewriting of the supercharges in terms of twisted variables can be repeated for the fermions of the theory and yields a set of $p$-forms $(\eta,\psi_\mu, \chi_{\mu\nu}, \ldots)$ which for the case of $Q=16$ matches the number of components of a real K\"ahler-Dirac field. This repackaging of the fermions of the theory into a K\"ahler-Dirac field is at the heart of how the discrete theory avoids fermion doubling as was shown by Becher, Joos and Rabin in the early days of lattice gauge theory \cite{Rabin:1981qj, Becher:1982ud}. 

The twisted four-dimensional ${\cal N} = 4$ theory can be written in a compact form involving five-dimensional fields \cite{Catterall:2009it}
\beq
\label{4daction}
S = \frac{1}{g^2} \int\Tr \cQ \left(\chi_{ab}\cF_{ab}+\eta [ \cDb_a,\cD_a ]-\frac{1}{2}\eta d\right) - \frac{1}{8g^2} \int \Tr \epsilon_{abcde} \chi_{de} \cDb_{c} \chi_{ab}~.
\eeq
The fields are now in $SU(5)$ representations with the Roman indices $a, b$ running from $1, \cdots, 5$. The sixteen fermionic degrees of freedom are packed in the $SU(5)$ decomposition $(1 \oplus 5 \oplus \overline{10}) \rightarrow (\eta,\psi_a,\chi_{ab})$. The ten bosonic degrees of freedom are encoded in a complex gauge fields $\cA_a$ and $\cAb_a$, decomposing as $5 \oplus \overline{5}$ under $SU(5)$.  We have $\cA_a = A_a + iB_a$ with corresponding complexified field strength $\cF_{ab}$. The field $d$ is an auxiliary field introduced for the off-shell completion of the twisted algebra.

The nilpotent transformations associated with $\cQ$ are given explicitly by
\bec
  \begin{tabular}{  l  l  }
	$\cQ\; \cA_a = \psi_a,$ & $\cQ\; \chi_{ab}=-\cFb_{ab},$  \\ 
        $\cQ\; \psi_a = 0,$ & $\cQ\; \eta = d,$  \\ 
	$\cQ\; \cAb_a =0,$ & $\cQ\; d = 0.$
  \end{tabular}
\eec
The action (\ref{4daction}), after $\cQ$-variation, integrating out the auxiliary field $d$ and dimensionally reducing along the 5th direction gives the four-dimensional action of the twisted ${\cal N} = 4$ theory
\bea
S &=& \frac{1}{g^2}\int \Tr \Big(\cFb_{\mu \nu} \cF_{\mu \nu} + \frac{1}{2}{[}\cDb_{\mu}, \cD_{\mu}]^2 + \frac{1}{2}{[}\phib, \phi]^2 + (\cD_{\mu}\phi)^{\dagger} (\cD_{\mu}\phi) - \chi_{\mu \nu} \cD_{[\mu} \psi_{\nu]} \nn \\
\label{eq:susy-lattice-twist-action}
&& -\psib_{\mu} \cD_{\mu} \etab - \psib [\phi, \psi_{\mu}] - \eta \cDb_{\mu} \psi_{\mu} - \eta {[}\phib, \etab] - \chi^*_{\mu \nu} \cDb_{\mu} \psib_{\nu} - \chi^*_{\mu \nu} {[}\phib, \chi_{\mu \nu}]\Big),
\eea
where the Greek indices run from $1, \cdots, 4$; the last two terms arise from the dimensional reduction of the $\cQ$-closed term; and $\chi^*$ is the Hodge dual of $\chi$: $\chi^*_{\mu \nu} = \frac{1}{2} \epsilon_{\mu \nu \rho \lambda} \chi_{\rho \lambda}$. (See \cite{Catterall:2009it} for more details.) 

This theory can now be further dimensionally reduced to two dimensions.
\subsection{Discretization of the twisted theory}
We follow the geometric discretization scheme to write down the lattice version of the theory. The complex gauge fields are represented as complexified Wilson gauge links $\cU_\mu(x)=e^{\cA_{\mu}}(x)$ living on links of a lattice which for the moment we can think of a square lattice. These link fields transform in the usual way under $U(N)$ lattice gauge transformations
\beq
\cU_\mu(x)\to G(x)\cU_\mu(x)G^\dagger(x)~.
\eeq
Supersymmetric invariance then implies that $\psi_\mu(x)$ live on the same links and transform identically. The scalar fermion $\eta(x)$ transforms as a site field: $\eta(x)\to G(x)\eta(x)G^\dagger(x)$. The field $\chi_{\mu\nu}$ is a 2-form and we choose it to lie along a diagonal link.

The continuum derivatives are replaced by difference operators on the lattice. We use $\cD^{(+)}_\mu$ for curl-like operations and $\cD^{(-)}_\mu$ for divergence-like operations. (For more details on the lattice implementation we refer the reader to \cite{Catterall:2009it}.) 

\section{Simulation results}
We have simulated the two-dimensional thermal theory in the range of inverse temperature $0.02 < \beta < 1.0$. We have focused on two observables, the spatial and temporal Polyakov lines. For a lattice of size $L \times T$ these are given respectively by
\beq
P_x = \frac{1}{N} \Big{\langle} \Big|\Tr \Pi_{a_x=0}^{L-1} U_{a_x}\Big| \Big{\rangle},~~P_t = \frac{1}{N} \Big{\langle} \Big|\Tr \Pi_{a_t=0}^{T-1} U_{a_t}\Big| \Big{\rangle}~.
\eeq

We have performed the simulations on a $2 \times 8$ lattice for the values of color $N=2, 3$ and $4$, with values of infrared regulator m = 0.05, 0.10 and 0.20. (To regulate the divergence with Monte Carlo simulations we have introduced a scalar mass term m.) The spatial Polyakov line undergoes a sudden transition from non-zero to zero values around $\beta \approx 0.2$ indicating that there is a phase transition in the gravitational theory. The temporal Polyakov loop does not undergo a transition indicating the presence of event horizons in the black solutions of the gravitational theory. We summarize the results in figures 1 and 2. More detailed results are given in \cite{sc-aj-tw}. 
\subsection{Acknowledgments}
SC and AJ are supported in part by the US Department of Energy under grant DE-FG02-85ER40237. TW is supported by a STFC advanced fellowship and a Halliday award. Simulations were performed using USQCD resources at Fermilab.
\begin{figure}[h]
\begin{minipage}{18pc}
\includegraphics[width=.7\textwidth, angle=270]{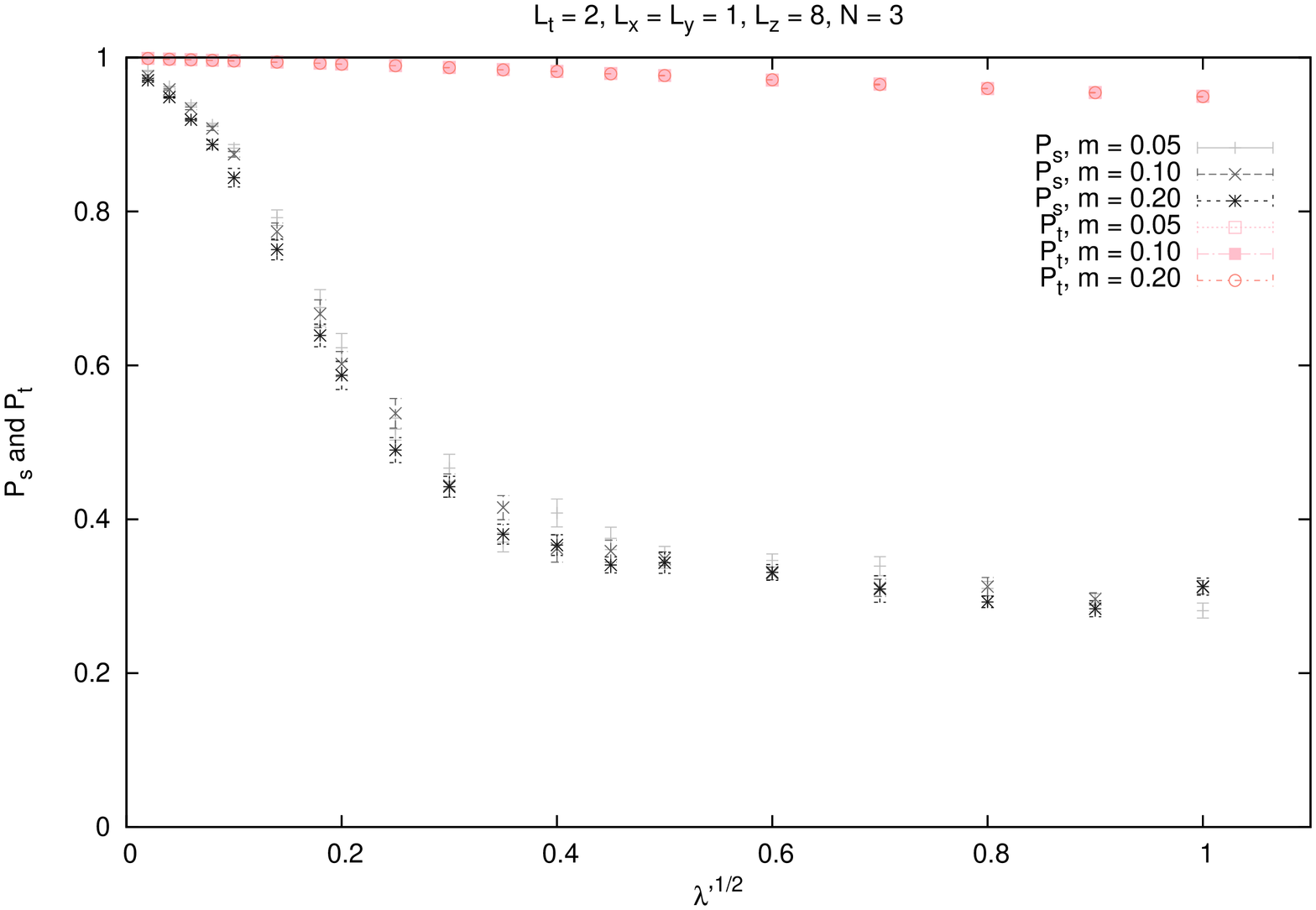}
\caption{\label{label1} \small{Spatial and temporal Polyakov lines ($P_s$ and $P_t$) vs square root of the coupling ($\sqrt{\lambda'}$) for maximally supersymmetric $SU(3)$ Yang-Mills on a $2 \times 8$ lattice using different values of the infrared regulator m.}}
\end{minipage}\hspace{2pc}%
\begin{minipage}{18pc}
\includegraphics[width=.7\textwidth, angle=270]{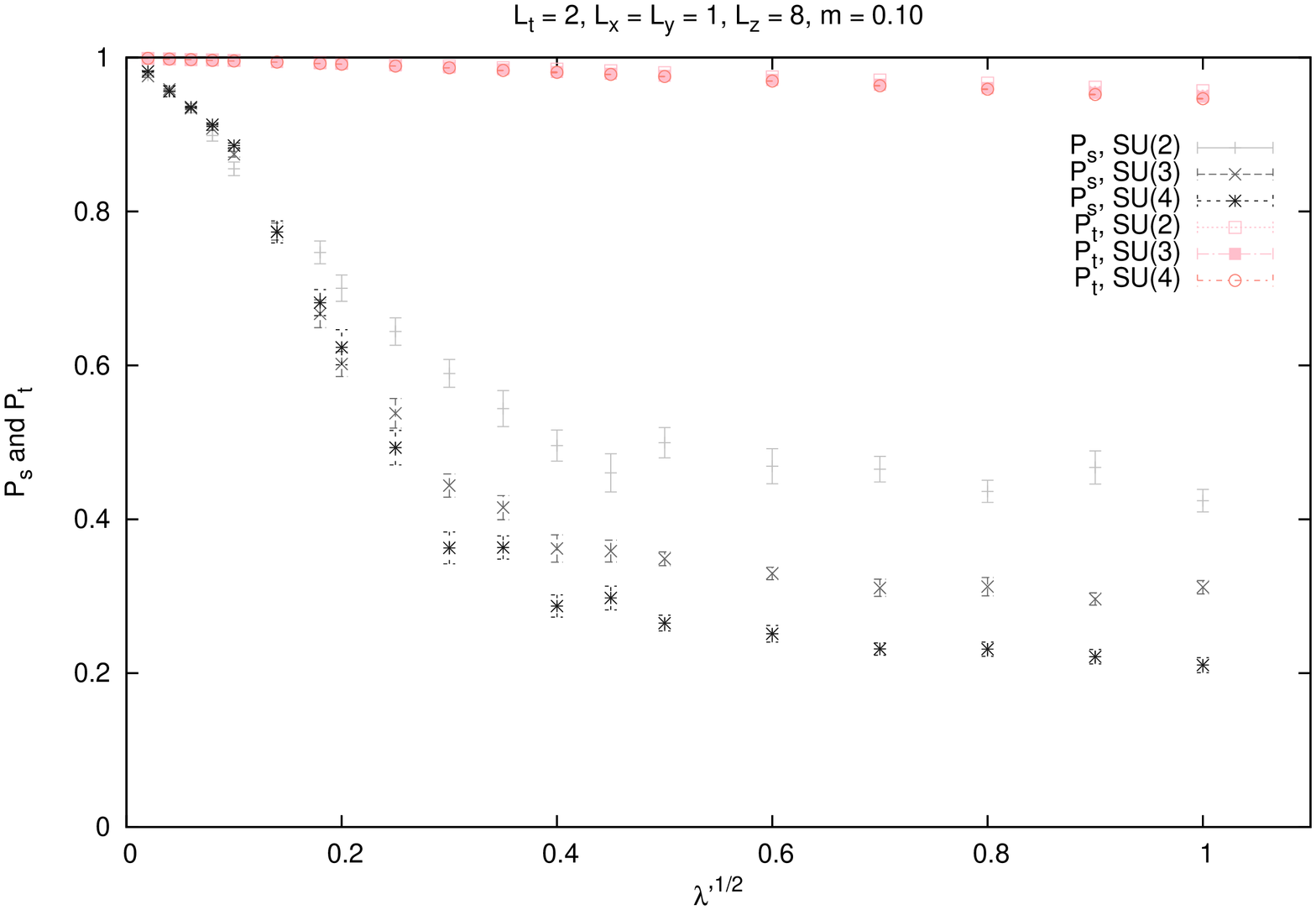}
\caption{\label{label2} \small{Spatial and temporal Polyakov lines ($P_s$ and $P_t$) vs square root of the coupling ($\sqrt{\lambda'}$) for maximally supersymmetric $SU(N)$ Yang-Mills with $N =2, 3, 4$ on a $2 \times 8$ lattice using the value m = 0.10 for the infrared regulator.}}
\end{minipage} 
\end{figure}

\end{document}